\documentstyle[12pt]{article}
\global\arraycolsep=2pt 

\newcommand{\bq}{\begin{eqnarray}}
\newcommand{\eq}{\end{eqnarray}}
\newcommand{\la}{\langle}
\newcommand{\ra}{\rangle}
\newcommand{\be}{\begin{equation}}
\newcommand{\ee}{\end{equation}}
\newcommand{\rarrow}{\rightarrow}

\begin{document}

\begin{titlepage}
\begin{flushright}
OUTP-95-25 P \\
hep-ph/9509322 \\
September 1995
\end{flushright}

\vspace{0.3cm}
\begin{center}
\Large\bf  
 {Two dimensional QCD with   matter
in adjoint representation: \\
What does it teach us?}\\
   \end{center}

\vspace {0.3cm}

 \begin{center} 
{\bf Ian I. Kogan\footnote{ 
On  leave of absence
from ITEP,  Moscow, Russia.}
\begin{center}
{\it  Theoretical
 Physics,
 1 Keble Road, Oxford, OX$13$NP, UK.}
\end{center}
  and \\
Ariel R. Zhitnitsky\footnote{
 e-mail address:arz@physics.ubc.ca  }}
 \end{center}
 
\begin{center}
{\it Physics Department, 
University of British Columbia, \\
6224 Agricultural Road, Vancouver,
BC V6T 1Z1, Canada.}
\end{center}
 \begin{abstract}

We analyse the highly excited states in 
$QCD_2 (N_{c}\rightarrow\infty) $  with 
 adjoint matter by using such general methods as 
dispersion relations, duality  and unitarity. 
We find the Hagedorn-like spectrum  
 $\rho(m) \sim m^{-a}\exp(\beta_H m)$ 
 where parameters $\beta_H$ and $a$ can be 
expressed in terms of 
 asymptotics of the following matrix elements $f_{n_{\{k\}}} \sim
 \la 0|Tr(\bar{\Psi}\Psi)^{k}|n_{k}\ra$. We argue that
    the asymptotical values $f_{n_{\{k\}}}$ 
 do not depend on $k$ (after appropriate normalization).
 Thus, we obtain  $\beta_H=
(2/\pi)\sqrt{\pi/g^2N_{c}}$  and $a = -3/2$
in case of Majorana  fermions in the  adjoint representation.
 The  Hagedorn temperature is the
 limiting temperature in this case.
 
We also argue that the chiral condensate 
$\la 0|Tr(\bar{\Psi}\Psi) |0\ra$ is not zero in the model.
 Contrary to  the 't Hooft model, this condensate
does not break down any continuous symmetries  and 
can not be considered as an order parameter. Thus, no Goldstone boson
appears as a consequence of the condensation.  

We also discuss a few   apparently different
 but actually  tightly related    problems: master field, condensate,
wee-partons and constituent quark model in the light cone framework.
\end{abstract} 
 
\newpage
\end{titlepage} 
\vskip 0.3cm
\noindent
{\bf 1. Introduction}
\vskip 0.3cm 

 Systems with a  density of  particle states
 increasing  exponentially with energy
\bq
 \rho(m) \sim m^{-a}\exp(\beta_{H} m)
\label{spectr}
\eq
 have been of significant interest  since  the early days of
the   statistical bootstrap and dual models \cite{hagedorn}.
  For such systems the
 canonical ensemble exists only for temperatures less
  than  the
 Hagedorn temperature $T_{H} = \beta_{H}^{-1}$, because for
 $\beta < \beta_{H}$ the partition function
 $Z = Tr \exp\left(-\beta H\right)$ is divergent. One can think 
 that  at $T_{H}$  one has  a
   confinement - deconfinement
 transition in four-dimensional gauge theories with a confinement 
 (hadron)  phase at $T < T_{H}$ and a  quark-gluon phase
 at $T> T_{H}$.

However one can ask whether   the Hagedorn temperature
 is the limiting one  or  there is a  phase transition in the
 system (\ref{spectr}). 
 As has been shown a long time ago by Frautschi and Carlitz
\cite{fracar}, who considered both canonical and microcanonical
 ensembles  for a gas of    weakly interacting  particles with
 a spectrum (\ref{spectr}), 
the answer to  this question   depends on the numerical value of
 the parameter $a$ in equation (\ref{spectr}) -
 for $a > (D+1)/2$ one has a phase transition with
 canonical and microcanonical  descriptions  being  not equivalent
 at high energies, while for
 $a \leq (D+1)/2$  one has $T_{H}$ as a limiting temperature
 and canonical and microcanonical  ensembles are  equivalent.
  Here  $D = d + 1$ is the  dimension of    space-time
 \footnote{Frautschi and Carlitz considered the case
 $D=4$.  The general case as well as $D = 10$ and/or $D= 26$ were  
considered
 in numerous papers in the mid-80's during the  
  (super)string   epoch. For review and references see,
 for example,  \cite{bowickreview}}. 

 To see what is the difference between $a$ larger and smaller
 then  $(D+1)/2$ let us consider the free energy of 
 noninteracting particles with the mass spectrum (\ref{spectr}).
 For bosons one gets
\bq
-\beta F_{b} = \ln Z_{b} = 
- \frac{V}{(2\pi)^{d}} \int_{m_{0}}^{\infty}
 dm \rho (m) \int d^{d} p \ln\left[ 1 - 
\exp\left(-\beta\sqrt{p^{2} + m^{2}}\right)\right]
 = \nonumber \\
=\frac{V}{(2\pi)^{d}} \int_{m_{0}}^{\infty}  dm \rho (m)
\sum_{n=1}^{\infty} \frac{1}{n} \int d^{d} p 
\exp\left(-n\beta\sqrt{p^{2} + m^{2}}\right)~~~~
\eq
 and for fermions
\bq
- \beta F_{f} = \ln Z_{f} = 
 \frac{V}{(2\pi)^{d}} \int_{m_{0}}^{\infty}
 dm \rho (m) \int d^{d} p \ln\left[ 1 + 
\exp\left(-\beta\sqrt{p^{2} +  m^{2}}\right)\right] =
 \nonumber \\
=\frac{V}{(2\pi)^{d}} \int_{m_{0}}^{\infty}  dm \rho (m)
\sum_{n=1}^{\infty} \frac{(-1)^{n+1}}{n} \int d^{d}p 
\exp\left(-n\beta\sqrt{p^{2} +  m^{2}}\right)~~~~
\eq
where $m_{0}$ is the infrared cut-off, which is usually of  the same  
order
 of magnitude as  $\beta_{H}$.
Because we are looking at  $\beta \sim \beta_{H}$   practically
  all particles  can be considered as non-relativistic ones and 
 we can   rewrite  
$\exp\left(-n\beta\sqrt{p^{2} + m^{2}}\right)$ 
as $\exp(-n\beta m-n\beta p^{2}/2m)
\left(1+O(n\beta p^{4}/m^{3})\right)$; 
it is easy to check  that the  neglected
 terms are of   order   $(n\beta m)^{-1}<<1$.
Integrating over $p$  we finally get
\bq
- \beta F_{b(f)}
=\frac{V}{(2\pi)^{d/2}\beta^{d/2}} \int_{m_{0}}^{\infty}  dm 
 m^{d/2}\rho (m)\sum_{n =1}^{\infty} 
\frac{[1(-1)]^{n+1}}{n^{d/2+1}} \exp(-n\beta m) 
\label{Fbf}
\eq
 Alternatively one can use the exact integral 
$\int d^{d}p \exp\left(-n\beta\sqrt{p^{2} + m^{2}}\right)
 \sim K_{D/2}(n\beta m)$ which 
 leads to (\ref{Fbf}) for $m >> \beta^{-1}$. 

It is clear that in the vicinity of the Hagedorn temperature
 $\beta \rarrow \beta_{H}$ 
  one can  neglect all terms in the sum with $n \geq 2$ (they will
 give singularities at $T = n T_{H}$)
 and  there is no difference between bosons and fermions in the
 leading singular term with $n=1$:
\bq
- \beta F =  \frac{V}{(2\pi\beta)^{d/2}}
\frac{1}{(\beta - \beta_{H})^{d/2 + 1 - a}}
\Gamma\left(\frac{D+1}{2}- a, (\beta-\beta_{H})m_{0}\right)\\ 
\nonumber
 + {\rm less~singular~terms} 
\label{frenerg}
\eq
where $\Gamma(x,y) = 
\int_{y}^{\infty} dt t^{x-1} e^{-t}$  is an incomplete
 $\Gamma$ function.

Now one can see that for  $ (D+1)/2 d/2 + 1 > a$ the 
  free energy  has a power singularity   when  $\beta \rarrow  
\beta_{H}$,
 whereas  in the case $ a > (D+1)/2$ it is  finite at 
 $\beta = \beta_{H}$.  For $a = (D+1)/2$ one has a logarithmic
 singularity $ \beta F \sim \ln\left((\beta-\beta_{H})m_{0}\right)$.
Thus in the case $ a < (D+1)/2$   $T_{H}$ is the limiting  
temperature -
 no matter how much energy we put into system its temperature will be
 less than $T_{H}$, contrary to the 
  case $ a > (D+1)/2$ where the energy density
 is finite at $ T = T_{H}$ and one has a phase transition point.

The only known examples  of spectra where the parameters $a$ are known
 analytically are the critical  bosonic and fermionic strings at
 $D = 26$ and $D = 10$ respectively. It is known 
(see \cite{bowickreview} for references) that in both cases one has
 $a_{closed} = D > (D+1)/2$ for closed and 
$a_{open} = (D-1)/2  < (D+1)/2$ for open strings, i.e. the Hagedorn
 temperature is 
 a limiting temperature for open strings.

 In this paper we shall   consider  the
 spectrum in a  $1+1$ dimensional   QCD gauge theory with 
   Majorana fermions in the adjoint representation
 with the action
\bq
S_{adj} = \int d^{2}x Tr \big[
-\frac{1}{4g^{2}} F_{\mu\nu}F^{\mu\nu} +
 i\bar{\Psi}\gamma^{\mu}D_{\mu}\Psi + m\bar{\Psi}\Psi\big]
\label{adjaction}
\eq
   The light-cone quantization
 of this  theory was considered in a 
 large $N_{c}$ limit in  \cite{adj}. The spectrum consists of 
 closed-string excitations. Contrary to the 't Hooft
 model  \cite{thooft}  with fermions in the fundamental
 representation of $SU(N_{c})$  describing the open-string
 excitations with only the  meson Regge trajectory, in this
 theory there is an infinite number of  closed-string
 Regge trajectories and the density of  particle states 
  is  of type (\ref{spectr})
  \cite{kutasov}, \cite{bdk}.
    The numerical value of $\beta_{H}
 \approx (0.7 - 0.75) \sqrt{\pi/(g^{2}N_{c})}$ 
   was found in the large $N_{c}$
 limit  in \cite{bdk}. The same  picture
 was obtained in   \cite{dkb} for a $1+1$- dimensional
 QCD with adjoint scalar quarks. The numerical value
 of the inverse Hagedorn temperature in this case is
  $\beta_{H}
 \approx (0.65 - 0.7) \sqrt{\pi/(g^{2}N_{c})}$. 
  Recent  progress in studying the spectrum in the light cone gauge
  was reported in \cite{Kutasov1}, \cite{AD}.
  However it was
 impossible to extract any information about the parameter $a$
  using  numerical methods.  
It is the main goal of this paper to calculate  both $\beta_{H}$
 and $a$ analytically using something like quark-hadron duality
 in a two-dimensional world. 

This  paper is organized as follows. In Sect.2 we use very 
general methods (like dispersion relations and duality) 
in order to extract  information  on highly excited states in
$QCD_2(N\rightarrow\infty)$ (exactly these states
determine the Hagedorn temperature).

In Sect.3 we   establish the connection between our results
(based on very general ideas) and  previous numerical calculations.
We argue that both results   agree  with each other  
 if  the  theory possesses the
  condensate $\la 0|Tr(\bar{\Psi}\Psi)|0\ra$.
 We argue that the nonzero magnitude for the condensate
is related somehow to the master field at large $N_c$.
However, an explicit realization of this connection is still lacking.
We also discuss the relation between the condensate 
$\la 0|Tr(\bar{\Psi}\Psi)|0\ra$ and the phenomenon of mixing
of components with different parton numbers.
The corresponding  analysis raises   another (but related) problem
about an infinite number
of constituents near zero momentum (so called wee partons).
We believe that all these problems are tightly connected to each other
and $QCD_2(N\rightarrow\infty)^{adj}$ is the perfect 
toy model to  elaborate these relations.

Sect.4 is our Conclusion and outlook.
 
\vskip 0.3cm
\noindent
{\bf 2. The spectrum of $QCD_2(N_c\rightarrow\infty$).}
 \vskip 0.3cm

First of all, we would like to recall an  example
of two dimensional QCD coupled to fundamental matter \cite{thooft},
  where  the  dispersion relations and duality, in turn, 
are very powerful tools.  Many interesting results
can be obtained by using these  methods.
The main idea is 
  to relate the known spectrum
of $QCD_2$ to the different vacuum characteristics. The same methods
are very useful for the study of  very general properties of the
 spectrum itself.

In particular, in the weak coupling regime,
$N_{c}\rightarrow\infty ,~g^2N_{c}\sim const. ,~g\ll  
m_q$
the chiral  condensate 
$\la\bar{\psi}\psi\ra $ at $m_q\rightarrow 0$
has been calculated {\it exactly} \cite{Zhit1}.
The original calculation was based on  
dispersion relations and duality.
The result was 
confirmed by numerical \cite{Ming1},\cite{Ming2} and 
independent analytical
calculations\cite{Lenz}. Moreover, the method has been  generalized for 
  nonzero quark mass and the corresponding explicit formula
for the chiral  condensate $\la \bar{q}q\ra$
with arbitrary $m_q$ has been obtained
\cite{Mattis}. Let us note that there is no contradiction with the Coleman 
theorem \cite{Coleman} at this point, because the BKT
(Berezinski-Kosterlitz-Thouless)\cite{BKT} behavior takes place in the
large $N$ limit\cite{Zhit1}. 

Furthemore, the so-called mixed vacuum condensates
with arbitrary number of gluon insertions, 
  $\la 0|\bar{q}(g\epsilon_{\mu\nu}
G_{\mu\nu})^nq |0\ra \sim M_{eff}^{2n}\la 0|\bar{q} q |0\ra  $,
 have also  been calculated \cite{CZ}. 
 We interpret the factorization property for the mixed vacuum condensates
 as   reminiscent of the master field at large $N$. 
 
Besides that, some  low-energy theorems can be obtained  
and from this we can 
 see  that there are no other 
states in addition to the  ones
found by 't Hooft. 
In other  words, if we miss some states  the dispersion 
and duality relations
  would indicate this.

Let us  demonstrate how it works in the simplest  case  
 and  consider the
asymptotic limit $ Q^2=-q^2\rightarrow\infty $
of the two-point correlation function \cite{ccg}, \cite{Zhit1}:
\begin{equation}
\label{1}
i\int dx e^{iqx}\la 0|T \{ \bar{\psi}i \gamma_5\psi(x),
\bar{\psi}i \gamma_5\psi(0) \} |0 \ra = \Pi(Q^2).
\end{equation}
It is clear, that the large $Q^2$ behavior of $\Pi(Q^2)$ is governed by
the free, massless theory, where
\begin{equation}
\label{2}
\Pi(Q^2\rightarrow\infty)=-\frac{N_{c}}{2\pi}\ln Q^2.
\end{equation}  
At the same time the dispersion relations state that
\begin{equation}
\label{3}
\Pi(Q^2)=\frac{N_{c}}{\pi}\sum_{n=0,2,4,...}\frac{f_n^2}{Q^2+m_n^2}
\end{equation}  
and the  sum is over states with an even number $n$
($n$ is the excitation number)  because
 we are considering  states with given parity. 
 Here the matrix elements  $f_n$ are defined as follows
\begin{equation}
\label{4}
\la 0|\bar{\psi}i \gamma_5\psi |n\ra=\sqrt{\frac{N_{c}}{\pi}}f_n  
,~~~n=0,2,4,.. . 
\end{equation}  
 Bearing in mind that for large  
$n,~~f_n^2 \rightarrow (g^2N_{c}/ \pi) \pi^{2}$
and $ m_n^2 \rightarrow (g^2N_{c}/\pi)\pi^2n $, we recover the
asymptotic result (\ref{2}). We can reverse the argument by saying that  
in order to reproduce $\ln Q^2$ dependence in (\ref{2}), the residues $f_n^2$  
must go to the definite constant $\sim (m_{n+1}^2-m_n^2)$ for large $n$
(this result can be considered as the strict constraint
of duality and the dispersion relations).
 
Now we want to repeat this analysis for the model we are interested
in, namely, for $QCD_2$ with adjoint matter.
As is known, the most important difference from t'Hooft model
is that the bound states may contain, in general, {\it any} number of  
quanta.
 In other words,  pair creation is not suppressed even in the
large $N$ limit. The problem becomes more complicated, but much more
interesting, because   pair creation imitates some physical
gluon effects.

We consider the following correlator analogous to (\ref{1}):
\begin{equation}
\label{5}
i\int dx e^{iqx}\la 0|T \{ \frac{1}{N_{c}}Tr\bar{\Psi} \Psi(x),
\frac{1}{N_{c}}Tr\bar{\Psi} \Psi(0) \} |0 \ra = P_2(Q^2),
\end{equation}
where $\bar{\Psi}= {\Psi}^T\gamma_0$ and label $P_2$ shows
the number of partons of external source $\bar{\Psi} \Psi(x)$;
the factor $1/N_{c}$ is included in  the definition
of the external current in order to make
the right hand side of the equation independent of  $N$.
 In the large $Q^{2}$ limit the   leading contribution to correlation
 functions is given by  the same diagram shown
 in  Fig. 1a). The result is
\begin{equation}
\label{6} 
P_2(Q^2\rightarrow\infty)=-\frac{2}{2\pi}(\frac{1}{2}\cdot 2)\ln Q^2. 
\end{equation}
The additional
factor $2$ in front of this expression 
comes from the two options in calculation of $Tr$ and
 is  related  to the $Z_{2}$ symmetry mentioned in \cite{kutasov}.
Besides that, we separate the trivial expression $\frac{1}{2}\cdot 2=1$ 
in parenthesis into  two parts.   The first factor
$\frac{1}{2}$ is related to the Majorana nature of the fermions
in comparison with Dirac's fermions in 't Hooft model.
The second factor $2$ which exactly compensates the first one,
is related to the normalization of $\Psi$ field ( we follow the
notation of  \cite{bdk}).

Now the problem arises. In the t'Hooft model we definitely knew
that only 2-particle bound states contribute to the corresponding
correlation function.
 This  is not true any more for the model under consideration and
any states,  in general,  may contribute to $P_2$. Without any small
 parameter at hand the task of  analyzing  the problem seems hopeless.
  However, one
 can see that in the large $Q$ limit  small parameters do arise.

 The {\it key} observation
is as follows: any pair creation effects are suppressed by a factor
$g^2N_{c}/Q^2$ because of the dimensionality of the coupling constant
in two dimensions  (in a big contrast with
real 4-dimensional QCD). Besides that, the quark mass term
produces the analogous small factor $m_q^2/Q^2$ and can be neglected
as well. Thus, the information about
highly excited states which provide the
$\ln Q^2$ dependence at large $Q^2$ can be obtained exclusively from
the analysis of the correlation function at large $Q^2$.
  In particular, the mass of highly excited
 bound states    should not depend on quark mass $m_q$
(it gives    small corrections to the leading
$\ln Q^2$ behavior).   With these remarks in mind  we suggest
the following pattern  which  saturates   the   correlation function
(\ref{5}, \ref{6}):
\begin{eqnarray}
\label{7}
\la 0|\frac{1}{N_{c}}\bar{\Psi}  \Psi |n_1\ra=
\sqrt{\frac{(m_0^2\pi^2)}{\pi}} f_{n_1} ,~~~n_1\gg 1,~~~~~
 n_1 \in 2Z   \nonumber  \\
m_{n_1}^2=m_0^2\pi^2n_1,~~f_{n_1}^2=1  
,~~m_0^2=\frac{2g^2N_{c}}{\pi}. 
\end{eqnarray}  
In this formula we assume  only one thing, namely 
that the  mass spectra for   massive states in $QCD_2$ coupled 
to fermions in the  adjoint and fundamental representations 
 are alike,  and  the  highly excited states look like those in the  
't Hooft model $m_n^2\sim n$. The dispersion relations state
in this case that the matrix elements (\ref{7}) do not depend
on excitation number $n$. Indeed,
 the only difference from the  't Hooft model is the doubling of the  
strength of
the interaction $g^2\rightarrow 2g^2$  \cite{bdk},  and the  
additional
degeneracy $Z_2$, mentioned above (\ref{6}).
The formula for  the  correlation function $P_2$ (\ref{6})
with the pattern (\ref{7}) can  easily be recovered:
\begin{equation}
\label{8} 
P_2(Q^2\rightarrow\infty)=
\frac{2}{\pi}\sum_{n_1=0,2,4...}\frac{(m_0^2\pi^2)f_{n_1}^2}{m_{n_1}^2 
+Q^2}
 \rightarrow \frac{2}{2\pi}\ln(\frac{\Lambda^2}{Q^2}),
\end{equation}
where $\Lambda^2$ is the ultraviolet cutoff and related to the  
subtraction
in the dispersion integral (insteadone can consider 
$P_2(Q^2)-P_2(0)$   which
is finite and does not depend on $\Lambda^2$ at all).
As before, any corrections to the asymptotic  expressions (\ref{7}),
like $ f_{n_1}^2=1  +0(1/n_1),~
m_{n_1}^2=m_0^2\pi^2n_1+ 0(m_q) $ produce some power corrections
$\sim 1/Q^2$ and they are not interesting at the moment.

 Thus  we see  the linear spectrum $m_n^2 \sim n$ unambiguously
 means that the   residues $f_{n_1} ,~n_1 \gg 1$  
do not depend on $n_1$, excitation number, for large $n_1$. This 
important consequence of the dispersion relations and duality
will be used heavily in our following discussions.
  
We would like to
 repeat this analysis for the  states with arbitrary number of
$\Psi(x)$ fields. To do so, let us introduce the currents which  
create  the $2k-$ parton states
(as will be discussed in the next section, the mixing between different
numbers of partons is not small. Thus, our term for the 
currents, creating the $2k-$ parton states, should be considered
as convention. ):
\begin{equation}
\label{9}
J_k(x)= \frac{1}{N_{c}^{k} }
Tr(\bar{\Psi}(x) \Psi(x))^k,~~~  
\la 0|J_k |~n_{\{k\}}\ra=\sqrt{\frac{(m_0^2\pi^2)}{\pi}}
\sqrt{\left(\frac{m_0^2\pi^2}{8\pi^2}\right)^{k-1}}~~f_{n_{\{k\}}}. 
\end{equation}
We introduced some numerical factors in  the right hand
side of  this formula for  future convenience. 
Let us  consider the following correlation function
  given  by the diagram of Fig.2:
\begin{equation}
\label{10}
i\int dx e^{iqx}\la 0|T \{ J_k(x),~
J_k(0) \} |0 \ra = P_{2k}(Q^2).
\end{equation}
As before we can calculate the asymptotic limit of
$P_{2k}(Q^2)$ at $Q^2\rightarrow\infty$, in which quark mass
 can be neglected, and, after taking into account only leading
 term in $1/N_{c}$ expansion, get the expression:
\bq
\label{11}
  P_{2k}(Q^2\rightarrow\infty)= 2k
 \int d^{2}x e^{iqx} \left[Tr\left(\frac{\hat{x}}{2\pi x^{2}}
\right)
 \left(\frac{\hat{x}}{2\pi x^{2}}
\right)\right]^{k} =  \nonumber \\
(-1)^k
 \frac{k}{\pi}\left(\frac{1}{8\pi^2}\right)^{k-1}
\frac{(Q^2)^{k-1}\ln Q^2}{ (k-1)!(k-1)!}
\eq
where $\hat{x} = x_{\mu}\gamma_{\mu}$.

 The $(Q^2)^{k-1}\ln Q^2/ (k-1)!(k-1)!$
 dependence in this equation is the  most important part. 
 We have to find the mass spectrum which will reproduce this
 large $Q^2$ behaviour. Of course one can not  define the
 pattern of the mass spectrum based only on the  dispersion integral,
 so we have to make some assumptions again. First of all we again
 make assumption  about the linearity of the spectrum which is
based on numerical results \cite{bdk}, \cite{dkb}. Using this assumption
  one can immediately  see  that  each mass level has to be
degenerate and our $k$-parton has to be classified by some 
 quantum numbers $n_1, n_2,...$ In principle one can imagine that
 for highly excited states the formfactors $f_{n_{\{k\}}}$ depend
 on the quantum number and in this case the dispersion integral
 will not provide sufficient information to calculate the
 asymptotic behaviour of the  spectrum. So we  make
 the second assumption that the constants $f_{n_{\{k\}}}$
  do not depend on $n_{\{k\}}$ in the limit
  $n_{\{k\}}\rightarrow\infty$. This assumption is close in
 spirit to the old ideas of ``hadron democracy''. It is hard to
 imagine any reasonable  pattern of strong dependence of
 formfactors $f_{n_{\{k\}}}$   on  quantum numbers $n_{\{k\}}$,
  and it is unclear  why some  excited
 states should be  produced more strongly than others, but for 
 now we have no any rigorous proof.
  In any case  these are  the assumptions we are making  here
 \footnote{ We also checked that these assumptions are selfconsistent with
 the large $Q^2$ asymptotics of some three- and four-point correlation
 functions of currents  under consideration.}. 

 Thus our highly excited  $k$-parton states
are classified by the $n_{\{k\}}=n_1, n_2,....n_k$ different numbers.
Furthermore, the dependence of $m_{n_{\{k\}}}^2 $ on $n_i$  is 
linear (one additional argument in favor of the linear spectrum
 is the equivalence  of the  massive sectors in
  $QCD_2^{adj}$ and  $ QCD_2^{fund}$ with $N_c$
 flavours , according to \cite{Kutasov1}).
The constants $f_{n_{\{k\}}}$
  do not depend on $n_{\{k\}}$ in the limit
  $n_{\{k\}}\rightarrow\infty$. 
It is clear, that these $n_{\{k\}}$ numbers should be identified
with the excitation numbers of the $ k $ partons.  Using 
the symmetry under:
 $ n_1 \leftrightarrow n_2 \leftrightarrow n_3 ...\leftrightarrow  
n_k$ 
we conclude that
$ m_{n_{\{k\}}}^2=m_0^2\pi^2(n_1+ n_2+...+n_k)$
 This formula was obtained previously \cite{kutasov}, but
we want to emphasize here that it has a much more general origin  which
is not related to the particular solution of the model suggested in
 \cite{kutasov}. 
The dispersion relations can now be written in the following form:
\bq
\label{12} 
P_{2k}(Q^2)=\frac{(m_0^2\pi^2)}{\pi}
\left(\frac{m_0^2\pi^2}{8\pi^2}\right)^{k-1}
\sum_{n_1,n_2,\ldots, n_k=0,2,4\ldots}\frac{f_{n_{\{k\}}}^2}
{\left(m_{n_{\{k\}}}^2+Q^2\right)}  \Rightarrow 
\eq
\bq
 \frac{(m_0^2\pi^2)}{\pi}
\left(\frac{m_0^2\pi^2}{8\pi^2}\right)^{k-1}
\sum_{N \gg 1}^{\infty}
\frac{f^{2}_{k}G_k(N)}{m_0^2\pi^2N+Q^2} =
(-1)^k \frac{k}{\pi}(\frac{1}{8\pi^2})^{k-1}
\frac{(Q^2)^{k-1}\ln Q^2}{(k-1)!(k-1)!} \nonumber 
\eq
where we introduced the notation $f_k$ for  the constants
which might depend on  the number of partons $k$,
but not on their excitation numbers $n_{\{k\}}$:
$f_{n_{\{k\}}}\rightarrow f_k,~~ n_{\{k\}}\rightarrow\infty$.
 We  also introduced
the factor of degeneracy $G_k(N=n_1+...n_k )$ 
for the bound states of $2k$ partons with the
given mass $m^2=m_0^2\pi^2N$   for highly  excited states.
Now, the only way to satisfy the  dispersion relation 
at asymptotically large $Q^2$ (after an
appropriate number of subtractions) is to
put
\begin{equation}
\label{13}
  G_k(N)=
 \frac{2k N^{k-1} }{(k-1)!(k-1)!} \frac{1}{f_{k}^{2}}, ~~N\gg 1,
\end{equation}
where the factor $2$ is related to the $Z_2$ symmetry mentioned above.
  In this case, by taking into account 
that  only the  even states contribute to (\ref{12}) 
  we exactly reproduce  eq.(\ref{12}).
 
 Having at hand the expression
for $G_k(N)$ it is very easy to calculate the total
number of excited $2k$-parton  states with mass less
than ${\bf  M^2}=m_0^2\pi^2 {\bf N},~~ {\bf N}\gg 1$:
\begin{equation}
\label{14}
  \sum_{N\sim 1}^{\bf N}G_k(N)\simeq
 \frac{2{\bf N}^k }{(k-1)!(k-1)!} \frac{1}{f_{k}^{2}}, ~~~{\bf N}\gg 1.
\end{equation}

 Our last step is the summation over all $2k$ bound states.
To perform such a calculation we need to make
some assumption about the  $k$-dependence of the matrix elements
$f_k$.   Such  information can be obtained scientifically
only from explicit dynamical calculations; note  that
the dispersion relations can constrain  the product
$f_k^2\cdot G_k(N)$, but not $f_k,~G_k$ separately.
However,   we can parametrize the $k$-dependence in 
the following general way:   
\bq
\label{fk}
  \frac{1}{f_{k}}
  = c^{k} k^{\alpha},~~~ n_{k} \rightarrow \infty
\eq
The parameter $c$ in this formula  comes
 from the normalization of the wave function and  a nonzero $\alpha$
 may arise because of  an interaction between different pairs of partons.
  Now the sum we are interested in  
\bq
 \int^{\bf M} \rho(m)dm
 \sim  
\sum_{k=1} k^{2\alpha+2}\frac{\left(c{\bf M}/m_{0}\pi\right)^{2k}}{k! k!}
\eq
can be estimated   by
using  a saddle point approximation where   the extremum of the sum is at  
 $k = \left(c{\bf M}/m_{0}\pi\right)$. After some algebra one gets
\begin{equation}
\label{final}
\rho(m) \sim  m^{2\alpha + 3/2} 
\exp\left[c\left(\frac{\sqrt{2}}{\pi}
\sqrt{\frac{\pi}{g^{2}N_{c}}}\right)
 m\right].
\end{equation}
This is the main result of our paper - we demonstrate that the
 Hagedorn spectrum is determined in terms of the asymptotics of the
 matrix elements $f_k$. The  parameters $c$ and $\alpha$ in (\ref{fk})
which describe these  asymptotics   
define the Hagedorn temperature $T_{H} \sim 1/c$ as well as parameter
 $a = -2\alpha - 3/2$.
   If $\alpha  > -3/2$  then
 $a = -2\alpha - 3/2   < (D+1)/2 = 3/2$  which means
 that the Hagedorn temperature is the limiting temperature of the system.

We would like to go further in order to make 
some specific assumption
  about the $k$ dependence (\ref{fk}).
The model under consideration is a  theory with Majorana fermions.
This  makes some difference in calculating the diagramm Fig.2;
 each extra loop gives an  extra factor $1/2$ in   comparison with 
 the Dirac case.
We believe that this extra factor comes into the definition of
 the matrix elements $f_k$.
Thus, we {\it assume} that the transition from Dirac
fermions to Majorana fermions gives  an  additional
factor $f_k^2\sim\frac{1}{2^k}$.
 Let us repeat again:
we can not prove or disprove  the appearance of this
additional element in   comparison with the case of 
Dirac fermions in  the 't Hooft model. It can be proven
only as the result of the dynamical calculations.
Nevertheless, assuming the existence of this additional
 factor   and  summing  over all $2k$ states 
in formula (\ref{14})we get
(we use the standard 
formula for  the  expansion of  Bessel function
$I_{0}(z) = \sum_{k=0}\frac{(z/2)^{2k}}{(k!)^2}$):
\begin{equation}
\label{17}
 \int^{\bf M} \rho(m)dm
 = 2 \frac{{\bf  M^2}}{m_0^2\pi^2}
\sum_{k=0} \frac{\left({\bf M}/m_{0}\pi 
 \right)^{2k}2^k}{(k)!(k)!}=  2 \frac{{\bf M}^2}{m_0^2\pi^2} I_0(z),
\end{equation}
 where
 $$
~z=\frac{2}{\pi}
\sqrt\frac{\pi}{g^2N_c} {\bf M}
$$
We expect that the corrections to this formula from  
the non-highly excited states as well as from mixing with  different
parton states might  give some power corrections (or even change
the order of the Bessel function $I_{\nu}$). However, this  does not effect
the asymptotic behaviour  $z\rightarrow\infty$, which is one and the same
for all Bessel functions 
$I_{\nu}(z) \rarrow  e^z /\sqrt{2\pi z}$ and gives us  the spectrum
\bq
\label{18}
\rho(m) \sim  m^{3/2} \exp\left(\beta_{H}m\right), ~~~~~
 \beta_{H} = \frac{2}{\pi}
\sqrt{\frac{\pi}{g^2N_c}} \approx 
0.64\sqrt\frac{\pi}{g^{2}N_{c}}
\eq
This formula is in agreement with the  general expression (\ref{final})
where $c=\sqrt{2},~\alpha=0$.
As we mentioned above, the parameter $\alpha$ describes
the interaction between different pairs of partons.
 We have no rigorous  arguments that $\alpha=0$.
However, for asymptotically high $n_k\gg 1$
it is very unlikely that the  interactions 
change the magnitude of $\alpha$.
    One type of argument is based on consideration of, let us say,
 a four-parton system (i.e. $k =2$) with large $n_{1}$ and $n_2$ but
 with $n_1 >> n_2$. The interaction between these two subsystems
 is suppressed  again by factor $\frac{N_c
g^2}{Q^2}$. Thus we do not
expect that
the interaction will 
change the asymptotic behavior of the correlation function.
Therefore, it can not contribute to $\alpha$.
   However these arguments  may be spoiled by some
   complicated behavior of the multiparton wave function
 $\Phi(x_1,\ldots,x_{2k})$ at small $x$.

Let us note that the  analytical result
(\ref{18}) for the inverse Hagedorn temperature is in 
  fair agreement with the numerical results \cite{bdk},
where the   value $(0.7\div 0.75)$ has been obtained 
(compare to our factor $\frac{2}{\pi}\simeq 0.64$).

\vskip 0.3cm
\noindent
{\bf 3. Condensate $\la 0|Tr(\bar{\Psi}\Psi) |0\ra$.}
 \vskip 0.3cm 

We already mentioned 
that we have a qualitative agreement with the numerical
results \cite{bdk} regarding the  Hagedorn temperature.
Another qualitative numerical result which
was mentioned in \cite{bdk} is that the wave function of a 
typical excited state
is a complicated mixture
of components with different parton numbers.

Apparently, such a mixing is very difficult to explain
by analysing the 
 correlation functions at large $Q^2$ (the method we 
use throughout  this paper).
Indeed, if we consider the non-diagonal correlator analogous to (\ref{5}),
but with different number of constituents ($2\rightarrow 4$, as example):
\begin{equation}
\label{19}
i\int dx e^{iqx}\la 0|T \{ \frac{1}{N_{c}^2}Tr(\bar{\Psi} \Psi(x))^2,~
\frac{1}{N_{c}}Tr\bar{\Psi} \Psi(0) \} |0 \ra = P_{2\rightarrow 4}(Q^2),
\end{equation}
one could naively think that this correlation function
is strongly suppressed  because of extra powers of either $g^2/Q^2$ 
 or  $m_q^2/Q^2$ ( see Fig.3).
The naive interpretation in this case would be 
that the highly excited states  saturating 
an appropriate  dispersion relation  are pure
states with definite number of partons. Any mixing
$ 2 ~partons \rightarrow 4~partons$
  is highly
suppressed. Such a conclusion is in   severe contradiction
with  the numerical results \cite{bdk}. What is wrong?

We see only  one  (but very natural) resolution 
of this puzzle. Namely, we {\bf assume} that
 the theory possesses the 
condensate 
\be
\label{condensate}
\la 0 |Tr(\bar{\Psi} \Psi(x))|0\ra =\mu N_c^2\neq 0 .
\ee
In that case the suppression will be gone 
(here $\mu$ is some dimensional number,
   proportional  in the limit $m_q\rightarrow 0$  to 
  $m_0$.).
Indeed, the asymptotically
leading (at $Q^2\rightarrow\infty$) contribution  to eq.(\ref{19}) 
will be determined by the diagramm 
Fig.4  and not Fig.3. Thus, it is equal to
\bq 
\label{20}
P_{2\rightarrow 4}(Q^2)\sim
i\int dx e^{iqx}\la 0|T \{ \frac{1}{N_{c}^2}Tr(\bar{\Psi} \Psi(x))^2,~
\frac{1}{N_{c}}Tr\bar{\Psi} \Psi(0) \} |0 \ra  
\sim     \nonumber  \\
 \frac{\la 0|Tr\bar{\Psi} \Psi(0) \} |0 \ra }{N_c}\cdot
\frac{\delta^i_j}{N_c}\cdot
i\int dx e^{iqx}
\la 0|T \{ \frac{1}{N_{c} } \bar{\Psi}^k_i \Psi_k^j (x),~
\frac{1}{N_{c}}Tr\bar{\Psi} \Psi(0) \} |0 \ra . 
\eq

The right hand side of this expression coincides with  
 the correlation  function (\ref{5}) and  is not suppressed.
Once again, we have assumed that the chiral condensate
is not zero in this theory. Having made this assumption, 
we  can explain  theoretically  the numerical results \cite{bdk}
on strong mixing of components with different parton numbers.

Formula (\ref{20}) suggests the following
pattern of saturation of the corresponding dispersion relation.
Let us    denote  the state $ |X\ra$ as  the eigenstate with mass $m_X$
which contributes to  the asymptotic behavior of  
  the correlation function (\ref{5})(these states are not necessary 
two particle states).
The corresponding matrix element will be  denoted as $f_X$:
\be
\label{21}
\la 0|\frac{1}{N_{c}}Tr(\bar{\Psi}  \Psi) |X\ra=f_X.
\ee
Now, from the asymptotic expression (\ref{20})
one can see that the dispersion relation
corresponding to correlator (\ref{20}) will be {\bf automatically
satisfied} if the appropriate matrix element
can be expressed in terms of $f_X$ in the following way:
\be
\label{22}
\la 0|\frac{Tr(\bar{\Psi}  \Psi)^2 }{N_{c}^2}|X\ra\sim
\la 0|\frac{Tr(\bar{\Psi}  \Psi) }{N_{c}}|X\ra 
 \frac{\la 0|Tr(\bar{\Psi} \Psi(0))  |0 \ra }{N_c^2}
\sim f_X \frac{\la 0|Tr(\bar{\Psi} \Psi(0))|0 \ra }{N_c^2}
 \ee

An analogous expression can be written for an arbitrary
non-diagonal correlator. Saturation 
of the corresponding dispersion relation 
will  occur  automatically  for all such  cases.
The reason is that the calculation of a nondiagonal correlation function
(like (\ref{20})) using the  factorization
of the $\la 0|Tr(\bar{\Psi} \Psi(0))|0 \ra $ term  (see Fig.4)
and the evaluation of the matrix element (like(\ref{22})),
by factorizing
the same expression $\la 0|Tr(\bar{\Psi} \Psi(0))|0 \ra $, 
is {\bf one and the same } procedure. 

The moral of this discussion is very simple: in order to have
a mixing (to be in agreement with
the numerical calculations) 
 we have to have  some   non-suppressed off-diagonal
correlation functions. This goal can be achieved only by assuming
a non-zero magnitude for the condensate 
$\la 0|Tr(\bar{\Psi} \Psi(0))|0 \ra $.
In  other  words: the non-zero magnitude for the chiral condensate
and the phenomenon of mixing of components
with different parton numbers are  two sides of the same
coin. The dispersion relations connect these
two, apparently different, phenomena.

In what follows we give some  arguments that such a condensation
is a very natural phenomena in the theory under consideration
and it is very likely to happen.
To avoid any 
confusion, we would like to note from the very
beginning of these discussions, that this condensate
does not break down any continuous symmetry; thus, no goldstone boson
appears as a consequence of the condensation.

It has been known for a long time \cite{hooft1}  that
  gauge  theory    with 
the matter in the adjoint representation 
has unique topological properties.
Namely, in contrast with the standard QCD, 
the gauge group $G$ is $SU(N)/Z_N$ rather than just $SU(N)$
\footnote{$Z_N$ in this formula represents the elements of the center
 of the group.
As is known, the adjoint fields are not transformed
under the action of the elements of the
center.}. In the formal terms it means that the first
homotopy group $$\pi_1(\frac{SU(N)}{Z_N})=Z_N$$ is not trivial,
so that there are $N$ topologically nonequivalent sectors.
This is a new  classification which 
goes together with the standard topological analysis.

These arguments are very general, and thus  they are not specific to
the space-time dimensionality of the theory.
In particular, the realization of this extra symmetry
in four dimensional Yang-Mills theory
leads to the so-called toron solutions \cite{hooft2}
which may be responsible for the chiral condensate
$\la 0|\lambda^2   |0 \ra $ in supersymmetric Yang-Mills
theory\cite{cohen},
\cite{arz}
\footnote{Let us 
recall that this model contains  the standard gluon fields
as well as the  adjoint matter fermions, so-called gluino fileds $\lambda$.}.
The vacuum condensate  $$\la 0|\lambda^2 
 |0 \ra_k\sim\exp(\frac{i2\pi k}{N}) $$ 
in this model labels the different vacuum states 
 marked by the integer number $k$.
 The physics in all these vacua is the same and the number of
these vacuum states is $N$ in agreement with the additional
classification mentioned above and the value of the  
Witten index which equals $N$\cite{witten2}.

 A  very similar phenomenon takes place in two dimensional
supersymmetric $CP^{N-1}$ models.
In this case, again, there is a condensate \cite{arz} 
which classifies $N$ different vacuum states 
$\la 0|\lambda^2 \} |0 \ra_k\sim\exp(\frac{i2\pi k}{N})$.

Therefore, it is natural  to  expect that   $QCD_2$ with
 the gauge group $SU(N)$ and   matter in the adjoint representation
 has a kind of discrete $$\theta=\frac{2\pi k}{N}$$ vacua. 
Indeed in the ref. \cite{witten1} it was shown 
that the non-trivial topological classification
related to 
$\pi_1(SU(N)/Z_N)=Z_N$
leads to $N$ different vacuum states.
Therefore  it is natural  to think that in the presence
of adjoint matter these vacuum states are classified
by the vacuum condensate. 
The recent   explicit calculations of the
condensate  for $SU(2)$ \cite{smilga} and $SU(3)$ \cite{shifman}
 gauge groups    support this conjecture.

Let us repeat again that  we do not attempt to  calculate the
condensate in the present paper. Instead we  make a
 conjecture about the existence of this condensate
based on the previous analysis of  the model.
In this case we are in a position to explain
the numerical results about the spectrum in the model \cite{bdk}.
We could reverse our arguments by saying that
the complicated mixing discovered in the numerical
calculation can be explained from the theoretical
point of view only if the theory possesses a  
chiral condensate.

We believe that the condensation of the 
fermion adjoint  field in the theory is a  very 
important issue. Thus, we would like to
present some independent arguments 
in favour of this conjecture.
 
The  existence of the  condensate
$\la 0|Tr\bar{\Psi} \Psi(0)  |0 \ra $ in $QCD_2(N_c)$ with
Majorana fermions can be seen using the   
bosonized version of the theory\cite{smilga} . As is known,
 there is a  one to one correspondence between the original 
Majorana  
fermion bilinear $ \bar{\Psi}^a \Psi^b,$
 and some other   boson field 
given by an orthogonal matrix $\Phi^{ab}$ \cite{witten}. The precise
correspondence
\be
\label{23}
\bar{\Psi}^a \Psi^b=\mu \Phi^{ab}, ~~~~a,b=1,2...N_c^2-1 
\ee
depends on the regularization  procedure
and normalization convention. If the standard normalization
 $\la 0|\Phi^{ab}|0\ra=\delta^{ab}$
is chosen, then $\la 0 |  \bar{\Psi}^a \Psi^a |0\ra =\mu N_c^2$
with  $\mu\sim m_0$.
Thus,   in the   bosonized framework  we should not be surprised
 by  appearance of a condensate (it is a very natural phenomenon).
Rather, we should be surprised  by  the fact
that sometimes (e.g. $QCD_2$ with fundamental fermions and  finite $N_c$)
the condensate is zero.

 To clarify the last statement and to demonstrate the
difference between Dirac and Majorana fermions,
let us consider $QCD_2(N_c)$ with fundamental fermions
for arbitrary $N_c$ (not necessary  the 't Hoft model
with  $N_c\rightarrow\infty$).
The standard bosonisation rules for color
singlet operators (only this   part is important
for the following discussions) have the form:
\be
\label{24}
\bar{\psi}(1\pm\gamma)\psi\rightarrow\exp(\mp i2\sqrt{\pi}\theta(x)),
~\bar{\psi}i\partial_{\nu} \gamma_{\nu} \psi\rightarrow
\frac{N_c}{2}(\partial_{\nu}\theta(x))^2 
\ee
Taking into account this formula one can   calculate
the following correlation function 
(which actually determines the condensate) at large distances\cite{Zhit1}:
\begin{equation}
\label{25}
 \la 0|T \{ \bar{\psi} (1+\gamma_5)\psi(x),
\bar{\psi}(1- \gamma_5)\psi(0) \} |0 \ra  \sim
\exp(i\frac{2\pi}{N_c}\Delta(x))\sim  x^{-\frac{1}{N_c}}.
\end{equation}
Here $i\pi\Delta(x)\sim -\ln(x^2)$ is the propagator
of the {\bf massless} scalar field $\theta(x)$. The fact
that the field $\theta(x)$ remains massless even  when
 interactions are  taken into
account  was crucial in the deriviation  of (\ref{25}).
This fact is clearly related to the chiral symmetry $\theta\rightarrow
\theta + const$
of the original Lagrangian and the absence of an anomaly
in the singlet channel, which could break this symmetry
explicitly (like in  the Schwinger model).
 Besides that, in deriving  (\ref{25}) we took into account that
in the infrared region ($x\rightarrow\infty$ ) only the term of lowest
dimension in the Lagrangian $\sim \frac{N_c}{2}(\partial_{\nu}\theta(x))^2 $
is important. 

This deriviation explicitly demonstrates the 
difference between Dirac and Majorana fermions: in the former case there is
a symmetry which can not be broken and it prevents the condensation;
in the latter case there is no symmetry which could 
prevent the Majorana field from condensation.
Formula (\ref{25}) also explains the behavior of the  't Hooft
model in the large $N_c\rightarrow\infty$ limit, where
the correlation function (\ref{25}) goes to a constant. 
Such a behavior together with the  cluster property at $x\rightarrow\infty$
 implies  the existence of the condensate at $N_c=\infty$ in agreement
with the explicit calculation \cite{Zhit1}. At the same time,
for any finite but large $N_c$ the correlator falls off very slowly
demonstrating the BKT- behavior \cite{BKT} with no  
 contradiction of the   Coleman theorem.
   
We believe that we have convinced  the  reader that the condensation
is an unavoidable feature of the model  
(we do not know  how to calculate it, but that  is a  different issue).
If this is correct, let us formulate the following
question: what kind of field configurations are responsible
for this condensation? The honest answer to  this question --
we do not know. However   the  
important property of the large 
$N_c$ limit is that       the expectation
value of a product of any invariant operators 
 reduces to their factorized values\cite{Witten1}.
 Thus, one could expect that the {\bf classical
master field} saturates this condensate. However, we do not know
 how  this  might be explicitly realized.  
Let us note that a  similar situation occurs  place
in  the 't Hooft model, where
the mixed   vacuum condensates
with arbitrary number of gluon
insertions    $\la 0|\bar{q}(g\epsilon_{\mu\nu}
G_{\mu\nu})^nq |0\ra \sim M_{eff}^{2n}\la 0|\bar{q} q |0\ra $ 
can be explicitly calculated \cite{CZ}. Again, one could interpret
 such a factorization property  
 as    reminiscent of the master field at large $N_c$. 

One more issue which is related to the condensation of 
 the Majorana field is as follows.
We argued  earlier that 
the non-zero result for a  non-diagonal correlation function
analogous  to  (\ref{19}) 
 is due to the non-zero condensate 
$\la 0|Tr\bar{\Psi} \Psi(0)  |0 \ra $. At the same time
such   a correlation function describes 
the {\bf mixing}  between different numbers of partons.
This mixing is order of one (not small!)  just because the 
corresponding correlation
function is not suppressed! The same is true
for an arbitrary external current 
with an  arbitrary number of $(\bar{\Psi} \Psi)^k $ in it.
Thus, we arrive at  the following conclusion:
The condensation of the fermion field is {\bf responsible }
for the complicated mixture of 
components with different parton numbers in hadronic states.

Furthermore, because of the factorization of 
$\la 0|Tr\bar{\Psi} \Psi(0)  |0 \ra $ 
in the course of the calculation of hadronic matrix elements
 one could expect 
that the probability to have an additional  quark pair
is not small (the  direct   consequence of 
such a calculation   is the absence of   any 
suppression if one adds  a few more partons   to the system).
 Thus, the momentum which is carryied  by a 
 given parton in a hadron, is getting smaller and smaller
when we inject (without suppression)  additional partons
\footnote{Actually our arguments, based on a correlation function
consideration,  can be  applied only
for highly excited states. Only 
those states saturate the dispersion integral.
Only for those states can  information
about the correlation function   be transformed into
 knowledge of hadronic states.}.

Thus, we arrive  at  the next conclusion:
The condensation of the fermion field leads
to the  Feynman-Bjorken picture of {\bf wee} partons
\cite{FB}. Such a connection  
between condensation of  
$\la 0|Tr\bar{\Psi} \Psi(0)  |0 \ra $
and an infinite number of constituents near zero momentum
has been known for a long time( see the recent   review talk 
\cite{Susskind} delivered at the  Light Cone meeting by 
one of the pioneers  of the subject). The only comment we would like to make
here is as follows.

The nonzero condensate means that only part of the
sea quark distribution gets Lorentz contracted when
 the hadron is boosted, while another component
looks the same in all frames.   Fig.4
explains this statement in terms
of correlation function. Dispersion relations
transform this statement into one  about  hadrons.
   In the  four dimensional
world  this  would mean that the hadron does not become
a ``pancake" as $Q\rightarrow\infty$.  
We believe that the model under consideration is a   
model where: \\
a). the light cone quantization is already performed;\\
b). the mixing is believed to be  large;\\
c). the condensate $\la 0|Tr\bar{\Psi} \Psi(0)  |0 \ra $ is not zero;\\
d). the  Feynman-Bjorken picture of {\bf wee} partons emerged.\\
Thus it is a perfect model to explore the connections, mentioned
above,   between these  apparently different problems.
 
\vskip 0.3cm
\noindent
{\bf 4. Conclusion.}
 \vskip 0.3cm 

Let us note that that  the behaviour of the system 
at finite temperature
 $T$ is equivalent in the euclidian path integral
 formalism to  the behaviour of the same system  on 
the cylinder 
  $S^{1} \times R^{1}$ with radius $R = (2\pi T)^{-1} = 
(\beta/2\pi)$.
  One can  consider now compact directions $S^{1}$ as
 a space directions  and $R^{1}$ as an euclidian time. 

Then our results mean that there is
 a  minimal radius $R_{c} = \beta_{H}/2\pi$ -
 however one can ask 
 what will happen if we  consider QCD with 
adjoint matter living  on  a circle $S^{1}$ with radius
 $R < R_{c}$. Let us note that we can not
 reach this small radius smoothly starting from large 
$R$ because the  energy will diverge as 
 $E(R) \sim 1/(R-R_{c})^{3+2\alpha}$.
  This is nothing but the Casimir energy
which   becomes  singular
 at $R_{c}$ and prevents  the circle from  being  squeezed further.

Now in   the case  when $R<R_{c}$ from the very beginning we 
  definitely
 have no hadron spectrum -  the properties of this phase
   are  obscure. It is known, however, that the  spontaneous
 breaking of $Z_{N}$ symmetry in this model is  an unusual one
 \cite{kogan} - contrary to common
 belief two different $Z_N$ phases can not coexist  in
 the same space.   We also would like to mention here that it may
 be that  two theories with the same massive spectra, like adjoint
 $QCD_2$ and $QCD_2$ with $N_c$ fundamental fermions 
 (see \cite{Kutasov1}) may have different high temperature 
(small compactification
 radius) properties - indeed the first theory has a $Z_{N}$ degeneracy
 of the ground state in the high-temperature phase, and the second does not.

It is interesting to understand  what type of string theory 
 will reproduce the  behaviour of   $1+1$  QCD at high
 temperatures and/or  small radii  $R$, in particular what will
 be the spectrum of the winding modes, which  becomes tachyonic 
 \cite{vortices} at the point of the phase transition.  Let us note
 that in this theory besides the winding modes corresponding to
  pure gluon noncontractible  loops 
 $ L =  \frac{1}{N_{c}}{\rm tr}~
 P~\exp\left(i\oint_S^{1}  A_{\mu} dx^{\mu}\right)$ there are 
 more complicated objects with quarks
\bq
 L(x_1,..,x_n) \sim    \nonumber
\eq
\bq
{\rm tr}\left( P~\Psi(x_1)
\exp\left(i\int_{x_1}^{x_2}  A_{\mu} dx^{\mu}\right)
\Psi(x_2)
\ldots
\Psi(x_n)
\exp\left(i\int_{x_n}^{x_1}  A_{\mu} dx^{\mu}\right)
\right)
\label{L1n}
\eq
where $x_1,\ldots, x_n$ are the points on a noncontractible
 contour $S_{1}$.
In principle $L(x_1,..,x_n)$ may decay into usual hadron and
 pure gluon winding mode $L$, however  it may be that the most 
unstable winding modes are of the type (\ref{L1n}) and further
 investigation is necessary to answer this and many other
  questions about the high $T$  or small $R$ phase of this
 theory.

In the short term there are a number of areas in which progress can be made.
First, a  numerical calculation of the condensate  can be done  
 in the same way as in  the t'Hooft model.
Besides that we believe that the information about the condensate
is hidden in the numerical calculation (already completed)
 \cite{bdk} and luckily
it can be extracted from the  data bank.
The last (but not the least) issue is the exploration of 
the following relations:

$Master field\Leftrightarrow\la 0|Tr\bar{\Psi} \Psi(0)  |0 \ra
\Leftrightarrow Mixed ~states \Leftrightarrow wee ~partons
\Leftrightarrow quark ~model \Leftrightarrow Light ~cone ~framework
\Leftrightarrow Zero ~modes.$

{\bf Acknowledgements.}

\noindent

\bigskip

 We  would like to  thank  S. Dalley, K. Demeterfi, I. Klebanov
 and D. Kutasov and especially  David Gross
  for  interesting  discussions.  
  I.I.K. would like to thank J. Wheater for 
careful  reading of the
  paper  and critical remarks and acknowledge the  support of the
 Oxford PPARC Particle Theory grant GR/J 21354 and  Balliol College
 as well as the hospitality of G. Semenoff and N.Weiss at UBC,
 where the last part of this work was completed.
 A.R.Z. would like to thank  the participants of the Light Cone Meeting,
Telluride, CO, August 1995, where this work was  presented.
{\renewcommand{\Large}{\normalsize}

 
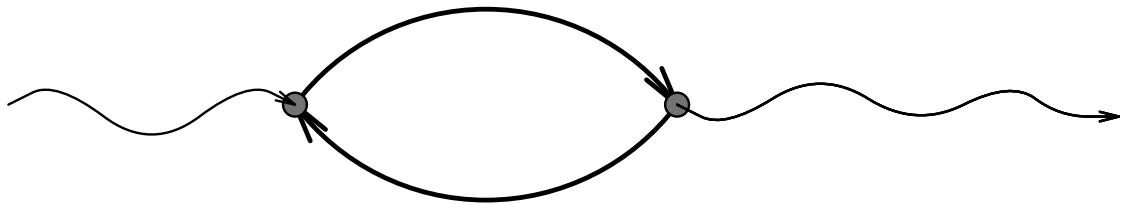
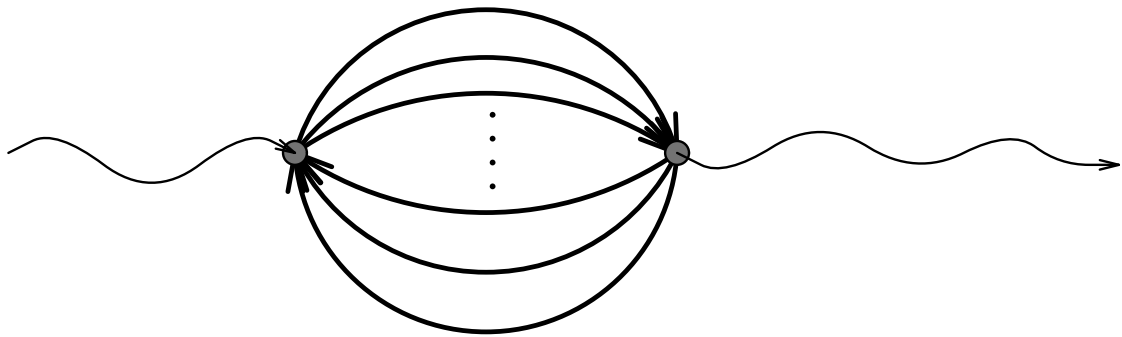
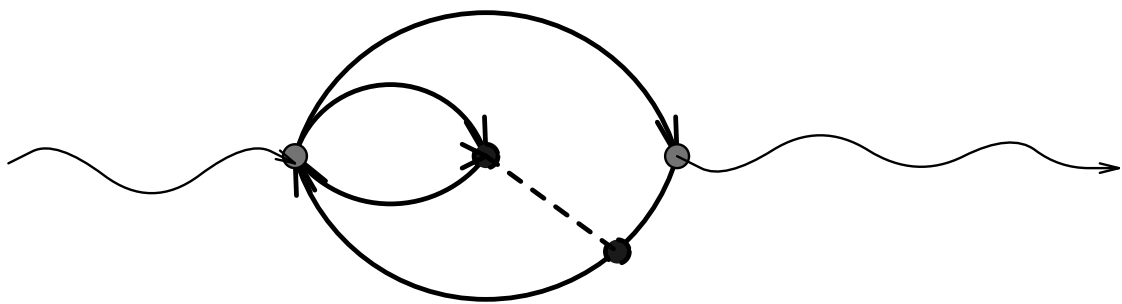
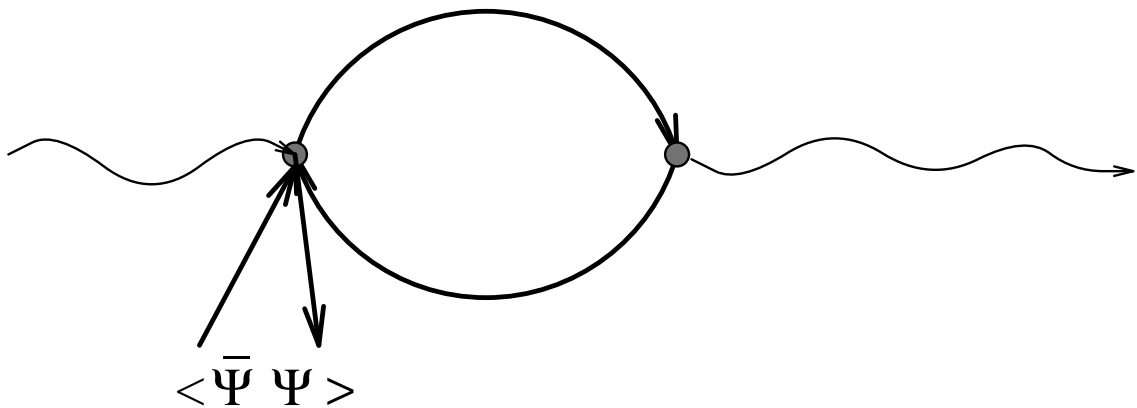
\end{document}